\documentclass[apj]{emulateapj}

\begin{document}

\title{The Coronal Structure of AB Doradus}

\author{O. Cohen\altaffilmark{1}, J.J. Drake\altaffilmark{1}, V.L. Kashyap\altaffilmark{1},
G.A.J. Hussain\altaffilmark{2}, T.I. Gombosi\altaffilmark{3}}

\altaffiltext{1}{Harvard-Smithsonian Center for Astrophysics, 60 Garden St. Cambridge, MA 02138}
\altaffiltext{2}{ESO, Karl-Schwarzschild-Str. 2, D-85748 Garching, Germany, Germany}
\altaffiltext{3}{Center for Space Environment Modeling, University of Michigan, 2455 Hayward St.,
Ann Arbor, MI 48109}

\begin{abstract}

We perform a numerical simulation of the corona of the young, rapidly rotating K0 dwarf AB Doradus 
using a global MHD model. The model is driven by a surface map of the radial magnetic field 
constructed using Zeeman-Doppler Imaging. We find that the global structure of the stellar corona is
dominated by strong azimuthal tangling of the magnetic field due to the rapid rotation. The MHD
solution enables us to calculate realistic Alfv\'en surfaces and we can therefore estimate the stellar 
mass loss rate and angular momentum loss rate without making undue theoretical simplifications. 
We consider three cases, parameterized by the base density of the corona, that span the range of 
possible solutions for the system. We find that overall, the mass and angular-momentum loss rates are 
higher than in the solar case; the mass loss rates are 10 to 500 times higher, and the angular momentum 
loss rate can be up to $3\times{10}^4$ higher than present day solar values. Our simulations show that
this model can be use to constrain the wide parameter space of stellar systems. It also shows that
an MHD approach can provide more information about the physical system over the commonly used
potential field extrapolation.

\end{abstract}



\section{INTRODUCTION}
\label{sec:Intro}

Rapidly rotating young stars are important for the study of stellar activity in two aspects.
First, these stars provide an ``enhanced'' picture of the fundamental stellar magnetic activity also 
seen in less active stars (such as our Sun) and therefore help to better understand the 
causes and consequences of stellar activity. Second, these rapidly rotating systems provide an
opportunity to observe the repeating rotationally modulated signatures of activity on relatively 
short time scales.

An example of a well observed young, active star is AB Doradus (HD 36705, AB Dor hereafter), 
a K0 dwarf with an age of about $75\;Myr$ \citep{Zuckerman04,luhman05,Nielsen05,LopezSantiago06,Janson07}. 
AB Dor spins with a rotation period of $P=0.5\;days$ \citep{Pakull81}, has a mass of about 
$M_\star=0.76M_\odot$ \citep{Guirado97}, and radius of about $R_\star=0.86R_\odot$ \citep{Maggio00} 
(slightly smaller than the solar radius, $R_\odot$, and solar mass $M_\odot$). AB Dor has been well 
observed over most of the electromagnetic spectrum 
\citep[e.g.][]{Lim94,Grothues97,Schmitt97,Vilhu98,Cutispoto98,Jarvine05,Budding09} and its activity 
has been studied in detail in the X-ray band 
\citep[e.g.][]{Maggio00,Gudel01,SanzForcada03,GarciaAlvarez05,Hussain05,Matranga05,Hussain07}. 
Most importantly for the work in hand, AB Dor has been extensively observed using Zeeman-Doppler 
Imaging (ZDI) \citep{DonatiSemel90,Donati09}. This method enables some inference of the vector 
magnetic field on the stellar surface based on the polarization of emitted light by the field and 
the modulation in Zeeman splitting as the star rotates. For rapidly rotating systems such as AB Dor, 
it is possible to use this method to reconstruct a large-scale map of the stellar surface magnetic 
topology, analogous to the solar synoptic ``magnetograms'' (27-day--average surface field distribution), 
albeit with lower spatial resolution. The ZDI technique has been used to map the surface magnetic field 
of AB Dor \citep{DonatiCollierCameron97,Donati99,Hussain02}, as well to study its stellar 
cyclic activity and surface differential rotation 
\citep{DonatiCollierCameron97,CollierCameron02,Pointer02,Jeffers07}. The stellar cycle of AB Dor 
has been also studied using Doppler Imaging observations by \cite{Jarvine05}.

It is commonly assumed that, like the solar corona, stellar coronae are dominated by their magnetic 
fields, so that the magnetic pressure, $P_B=B^2/8\pi$, is much greater than the thermal pressure, $P_{th}=nkT$, 
and that the plasma $\beta$ parameter, $\beta=P_{th}/P_B$, is much smaller than 1. In this case, the magnetic 
field can be assumed to be potential (i.e. there are no forces or currents acting on it) and it can be 
described as a gradient of a scalar potential. Under these assumptions, the three-dimensional distribution 
of the magnetic field can be obtained by solving Laplace's equation for the scalar potential, where the 
surface field maps are used as the inner boundary condition and the outer boundary condition assumes a 
purely radial field at a certain height above the surface (known as the ``source surface''). This 
technique of extrapolating the coronal magnetic field is known as the ``potential field'' method 
\citep{altschulernewkirk69,altschuler77}. It has been used extensively in solar studies and more recently 
to extrapolate the coronal magnetic fields of stars such as AB Dor 
\citep{Donati99,Jardine99,Hussain02,Jardine02,McIvor03}. In particular, \cite{Hussain07} have reconstructed 
the X-ray corona of AB Dor based on the potential field extrapolation, combined with different X-ray models 
for coronal loops.

The potential field extrapolation is a useful tool to obtain a first order approximation of the large-scale 
structure of a stellar corona based on its surface magnetic map. The approach taken by \cite{Hussain07} 
is justified due to the fact that the major part of the coronal X-ray emission is expected to originate 
from the smaller closed loops near the surface, which are usually in a near-potential state (in the static 
case where footpoint motions and other short-term motions are not taken into account). There are, however, 
good reasons to attempt a more physical approach for describing stellar coronae. First, the potential field 
approximation (by itself) provides information only about the magnetic field of the system, and does not 
address energy dissipation through driving a wind. Second, the location of the source surface is not well 
defined, and third, when considering a complete description of the physics involved, including conservation 
of mass, momentum, and energy, one needs to take into account the effects of coronal heating and stellar 
wind acceleration, the stretching of the field lines by the highly conductive coronal plasma to a 
non-potential state, as well as the effects of rapid rotation in stars like AB Dor.

Here we extend the work of \cite{Hussain07} and present a complete three-dimensional MagnetoHydroDynamic 
(MHD) simulation of the corona of AB Dor based on its observed surface magnetic field distribution. For 
our simulation, we use a global MHD model developed for the solar corona, which provides a self-consistent 
stellar wind solution driven by surface magnetic field maps. The end result is a steady state, MHD, 
non-potential solution of the corona and wind of AB Dor, which includes the distribution of the complete 
set of physical parameters in the simulation domain. This more complete solution provides a better 
understanding of the large-scale coronal structure. We highlight the differences between the coronae of 
young stars like AB Dor and the solar corona due to rapid rotation of the former. We also provide 
realistic calculations of the possible mass loss rates for AB Dor, parameterized by the coronal base 
density.

We present the numerical model and the observational constraints used in the simulation in 
Section~\ref{sec:Simulation}. The results are presented in Section~\ref{sec:Results}, and the main findings 
are discussed in Section~\ref{sec:Discussion}. We conclude this work in Section~\ref{sec:Conclusions}.


\section{NUMERICAL SIMULATION}
\label{sec:Simulation}

The simulation of AB Dor is done using the solar corona model by \citet{cohen07, cohen08b},
which is part of the Space Weather Modeling Framework (SWMF) \citep{toth05} and is based
on the generic MHD {\sc BATS-R-US} model by \citep{powell99}. The model is driven by surface
magnetic field maps, and the initial condition for the magnetic field, as well as the volumetric
energy input for the stellar wind acceleration, is based on the distribution of the potential field. 
In addition, the boundary condition for the surface plasma density, $\rho_0$, is scaled with the 
magnetic field so that the plasma at closed field regions is more dense than in open field regions, 
as observed for the solar case \citep{phillips95}.

In the solar case, the source surface is usually set to be at $r=2.5R_\odot$. In the case of AB Dor 
however, the surface distribution of the magnetic field contains large regions with strong field. 
Therefore, we expect loops on AB Dor to be much larger than solar loops so we 
choose to set the source surface at $r=10R_\star$. This should not have any effect on the non-potential, MHD 
solution since the potential field only serves at the initial condition. The MHD solution is mostly affected 
by the distribution of energy deposited into the stellar wind, and this energization is not sensitive to 
the location of the source surface as long as it is set above the the height of the largest closed loops. However, 
setting the source surface below the actual size of the loops (at $r=2.5R_\star$ for example), forces more 
field lines to be open and as a result, each plasma cell in the stellar wind is over-energized, resulting 
in solutions with unrealistically fast stellar winds.

A self-consistent wind acceleration in the code is obtained by assuming an empirical relation
between the magnetic flux tube expansion and the terminal stellar wind originating
from that flux tube. \cite{wangy90} and \cite{argepizzo00} have derived an empirical formula
that relates the final solar wind distribution, $u_{sw}$, to the flux tube expansion
factor, $f_s$. The factor $f_s$ is the ratio of the magnetic flux of a particular
flux tube at $r=R_{ss}$ and at $r=R_\star$, where $R_{ss}$ is the height of the source surface.
The empirical method described above predicts the spherical distribution of the solar wind
speed at $r\rightarrow \infty$. It is reasonable to assumed that far from the Sun (or star), 
the total energy equals to the bulk kinetic energy of the plasma, while on the solar surface, t
he total energy equals to the enthalpy of the fluid, minus the gravitational potential energy 
(the kinetic energy is zero). By adopting the conservation of total energy along a streamline 
(Bernoulli Integral), we can relate the final solar wind speed, $u_{sw}$ and the surface value of polytropic
index, $\gamma_0$, assuming the boundary conditions for the surface temperature, $T_0$, are known:
\begin{equation}
\frac{u^2_{sw}}{2}=\frac{\gamma_0}{\gamma_0-1}\frac{k_bT_0}{m_p}-\frac{GM_\star}{R_\star},
\label{BI}
\end{equation}
or
\begin{equation}
\gamma_0=\frac{\varepsilon}{\varepsilon-1},
\label{BI2}
\end{equation}
with
\begin{equation}
\varepsilon=\frac{m_p}{k_bT_0}\left( \frac{u^2_{sw}}{2}+\frac{GM_\star}{R_\star}\right).
\label{BI3}
\end{equation}
Here $k_b$ being the Boltzmann constant, $m_p$ the proton mass, and $G$ the gravitational constant.

Close to the Sun, the value of $\gamma$ is observed to be about unity,
(the plasma is highly turbulent), while at 1~AU $\gamma$ has a value
closer to 1.5 \citep{totten95,totten96}. This observed modulation in $\gamma$ can be related
to the powering of the solar wind, in the manner that the larger the gradient in $\gamma$
along a flux tube, the faster the wind flows along that tube. Based on this assumption,
and on the relation presented in Eq.~\ref{BI},
it is possible to construct a volumetric heating function, $E_\gamma(\gamma_0,\mathbf{r})$, 
in a way that the observed volumetric acceleration of the solar wind can be recovered. The 
additional term $E_\gamma \rightarrow 0$ as $\gamma \rightarrow 3/2$. 

The model described here constructs the particular spatial distribution of $E_\gamma$ based on the input
surface magnetic map and its potential field. It then solves the set of conservation laws
for mass, momentum, magnetic induction,and energy (the ideal MHD equations):
\newline
\begin{eqnarray}
&\frac{\partial \rho}{\partial t}+\nabla\cdot(\rho \mathbf{u})=0,& \nonumber \\
&\rho \frac{\partial \mathbf{u}}{\partial t}+
\nabla\cdot\left(
\rho\mathbf{u}\mathbf{u}+pI+\frac{B^2}{2\mu_0}I-\frac{\mathbf{B}\mathbf{B}}{\mu_0}
\right) = \rho\mathbf{g},& \nonumber \\
&\frac{\partial \mathbf{B}}{\partial t}+
\nabla\cdot(\mathbf{u}\mathbf{B}-\mathbf{B}\mathbf{u})=0, & \\
&\frac{\partial }{\partial t}\left(
\frac{1}{2}\rho u^2+\frac{1}{\Gamma-1}p+\frac{B^2}{2\mu_0}
\right)+ &
\nonumber \\
&
\nabla\cdot\left(
\frac{1}{2}\rho u^2\mathbf{u}+\frac{\Gamma}{\Gamma-1}p\mathbf{u}+
\frac{(\mathbf{B}\cdot\mathbf{B})\mathbf{u}-\mathbf{B}(\mathbf{B}\cdot\mathbf{u})}{\mu_0}
\right)=\rho(\mathbf{g}\cdot\mathbf{u})+E_\gamma, \nonumber &
\label{MHD}
\end{eqnarray}
with $\Gamma=3/2$ until a steady-state stellar wind solution
is obtained. The stellar input parameters required for the model are the boundary value for the density,
$\rho_0$, as well as the stellar radius, $R_\star$, mass, $M_\star$, and rotation frequency,
$\Omega_\star$. A summary of the stellar parameters of AB Dor adopted 
for the simulation, based on the references cited in Section~\ref{sec:Intro}, is provided in Table~\ref{table:t1}.

In the simulations presented here, we assume that the relation between the flux tube expansion and
the terminal speed obtained from that flux tube is a universal process that occurs on
AB Dor in a similar manner to the Sun. Observations of the corona of AB~Dor indicate dominant plasma 
temperatures peaking in the range 3--30~MK \citep[e.g.][]{SanzForcada03,GarciaAlvarez05}. $T_0$ is, 
in principle, the average temperature of the stellar corona. However, we stress that in 
our model, $T_0$ is essentially a free parameter for the boundary condition that controls the 
energization of the stellar wind. Further description of the adaptation of a solar corona model 
to stellar coronae can be found in \citep{Cohen10}.

\begin{figure*}[h!]
\centering
\includegraphics[width=6.in]{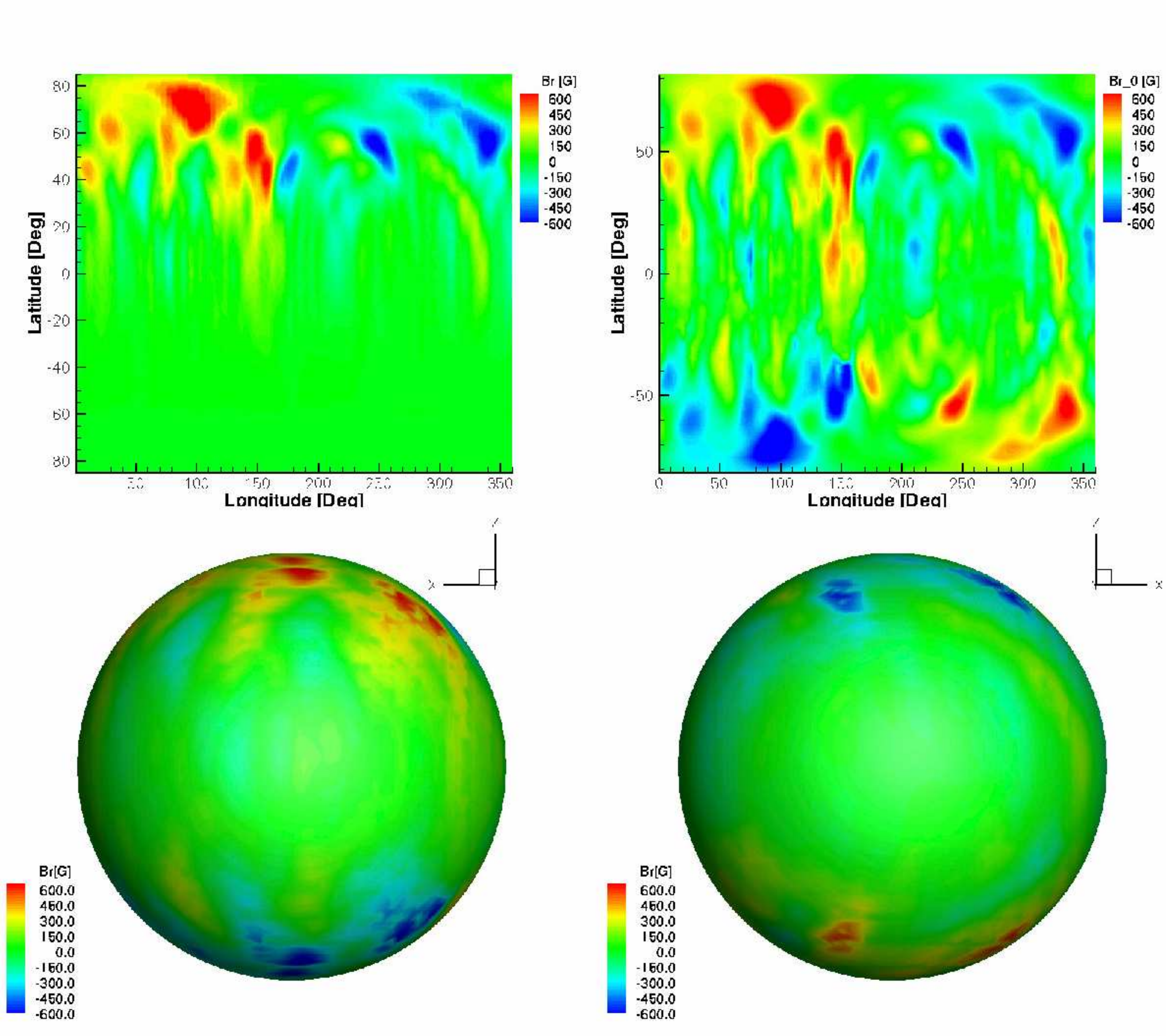}
\caption{Longitude-latitude map of the original, incomplete ZDI based map of AB Dor (top-left), and the interpolated map used in
the simulation displayed as a longitude-latitude map (top-right) and on two corresponding longitudinal spheres (bottom row).}
\label{fig:f1}
\end{figure*}

\begin{figure*}[h!]
\centering
\includegraphics[width=6.in]{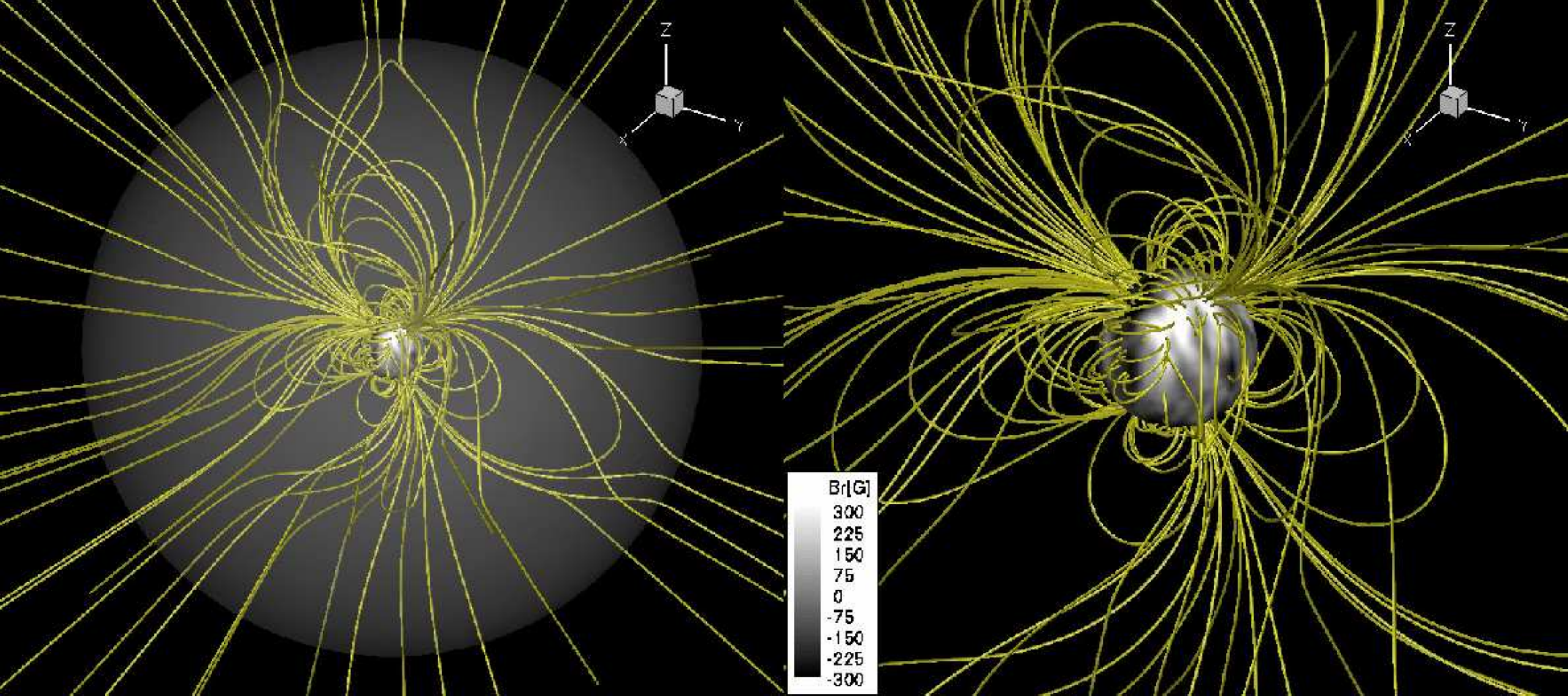}
\caption{The global structure of the potential field extrapolation for AB Dor (left). The small sphere in the middle 
represents the stellar surface
colored with grayscale contours of the radial field, while yellow lines represent the magnetic field lines. The outer white 
spherical shade represents the source surface located at $r=10R_\star$. Right panel shows a zoom close to the stellar surface.}
\label{fig:f2}
\end{figure*}

Figure~\ref{fig:f1} shows the the input surface magnetic field map adopted for the simulations. 
This is based on spectropolarimetric observations obtained in December 2007 (Hussain et al.
submitted to MNRAS), and analyzed in a similar 
manner to the maps described in \cite{Hussain02,Hussain05,Hussain07}. The reader is referred to those 
works for further details. Since AB Dor has an inclination of $60\deg$ \citep{Kuersteretal94}, the 
part of the stellar surface near the far pole is always hidden from view. Consequently, surface maps 
for AB~Dor are intrinsically incomplete. The initial, incomplete map is shown in the top left panel 
of Figure~\ref{fig:f1}. To construct a complete map, we enforced hemispherical reflection symmetry 
on the magnetic field across the equatorial plane. Those parts of the southern hemisphere with 
magnetic field magnitude of less than $50\;G$ were assigned magnetic field values from the same 
longitude at the corresponding northern hemisphere latitude, but with the opposite polarity. The 
complete map used in the simulation is shown as a longitude-latitude contour map (top-right panel) 
and as spherical plots of the two longitudinal hemispheres, colored with contours of the surface 
magnetic field (lower panels). Figure~\ref{fig:f2} shows the three-dimensional distribution of the 
potential field calculated based on this input surface map. It can be seen that the loops extend up 
to the height of the source surface (located at $r=10R_\star$) and that they have no toroidal component. 
The field is fully radial above the source surface as required by the analytical solution. 

We caution that there are a number of sources of systematic uncertainty present in the simulations. 
First, the ZDI maps are missing part of the stellar surface, and we have extrapolated the field, 
assuming antisymmetry, to those non-visible areas. Second, the magnetic field maps are of limited 
resolution (latitudinal resolution of 3 degrees) and do not resolve the details of the active regions. 
Third, large areas in the maps that appear to have strong fields may in fact be dominated by localized 
active regions that are smeared out due to the lower resolution; such a scenario would lead to a 
significantly different MHD solution. While the ZDI maps do likely miss low level structure, the maps 
should recover the strongest field regions, which are likely to dominate over the large-scale global 
models such as those considered here. The weaker complex fields would only be interesting much closer 
to the surface.

Observations of AB Dor reveal that the coronal base density, $n_0$, ranges between 
$10^{10}-10^{12}\;cm^{-3}$ \citep{SanzForcada03}. These measurements, however, were made based on measurements 
of strong emission lines that are associated with the denser, closed loops. Therefore, the density in the 
``quiet star'' (analogous to the ``quiet Sun'') where the wind originates should be lower. In order to 
partially cover this density range, we simulate three test cases with different values for the coronal base 
density. ``Case A'', with $n_0=2\cdot 10^8\;cm^{-3}$, which is the value used for simulations of the solar 
wind, ``Case B'' with $n_0=10^{9}\;cm^{-3}$, and ``Case C'' with $n_0=10^{10}\;cm^{-3}$. We simulate the wind 
and corona using a Cartesian box of $30R_\star$x$30R_\star$x$30R_\star$, in the frame of reference rotating 
with the star (to expedite convergence using the local time step algorithm \citep{cohen08b}). We use a 
non-uniform grid with a maximum resolution of $2\cdot10^{-2}R_\star$ prescribed near the surface. The grid 
is dynamically refined during the simulation so that high resolution is applied at location of magnetic 
field inversion (current sheets). We performed the simulation using the PLEIADES super computer at the 
NASA AMES center.


\section{RESULTS}
\label{sec:Results}

\begin{figure*}[h!]
\centering
\includegraphics[width=6.in]{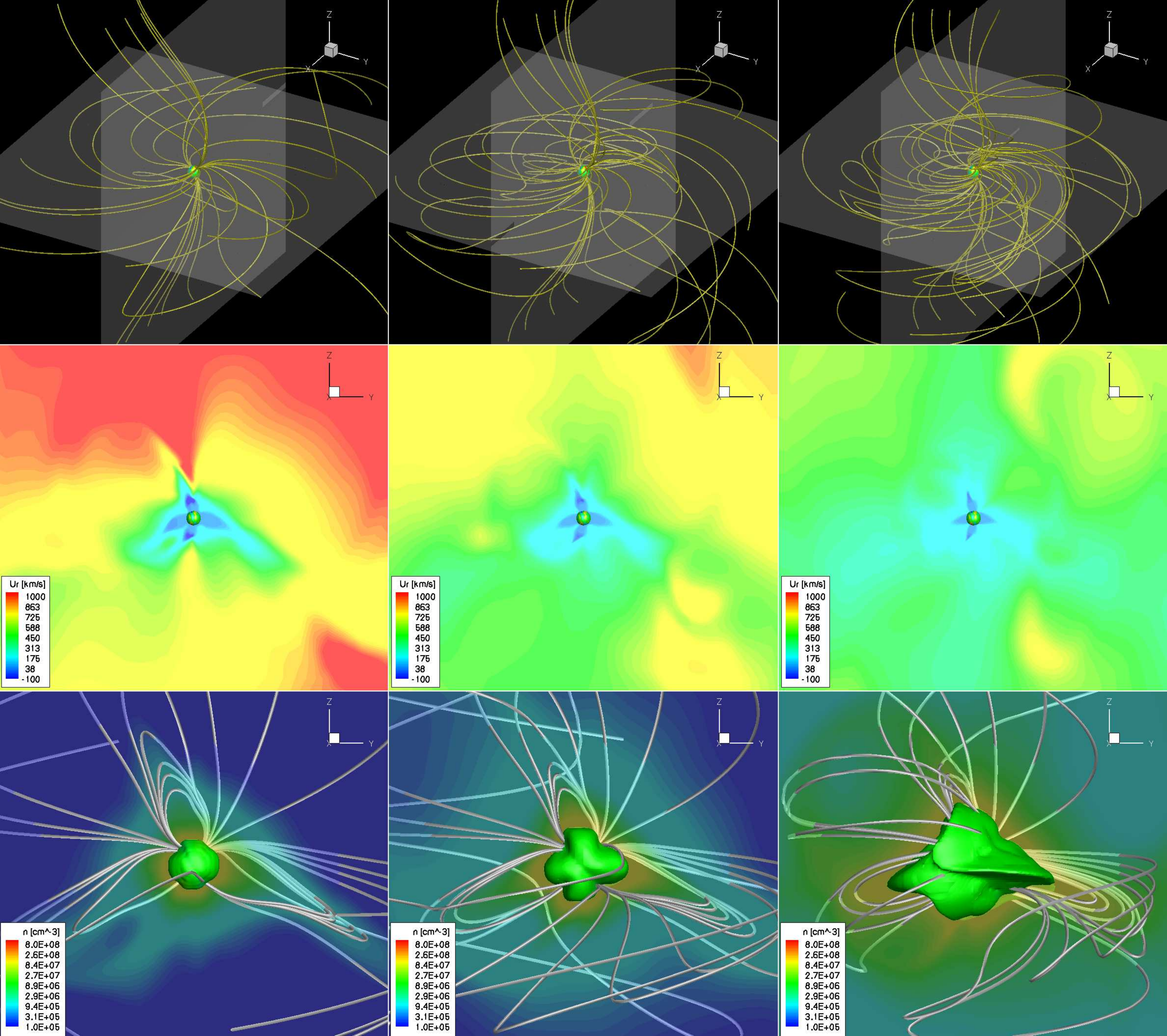}
\caption{Numerical MHD solutions of the coronal magnetic field (top), wind speed (middle), and plasma density (bottom) 
for AB Dor for cases A-C (left to right with increasing base density). Top: global structure of the MHD solution with $P_\star=0.5\;d$. The sphere 
represents the stellar surface colored with contours of the radial magnetic field, the yellow lines represent 
the three-dimensional magnetic field lines, and the $y=0$ and $z=0$ planes are shown as transparent white shades. 
Middle: A side view of the top panel without the field lines. Bottom: A side view of the top panel 
with color contours of number density and zoomed around the star. The green shade represents an iso surface of 
$n=10^8\; cm^{-3}$. Cases A-C are shown from left to right.}
\label{fig:f3}
\end{figure*}

The steady state MHD solutions for the three test cases are shown in Figure~\ref{fig:f3}.
The most notable feature of the solutions for all cases is the tangling of the field in the {\em azimuthal}
direction due to the rapid rotation of the star. This feature clearly does not appear in the potential
field solution, for which stellar rotation is not a relevant parameter, nor in similar MHD solutions for the Sun 
\citep[e.g.][]{cohen08b}.

As might be expected, solutions for the three cases are qualitatively quite similar, though closer inspection does
reveal significant differences. It can be seen from the middle panel of Figure~\ref{fig:f3} that the radial wind
speed decreases with an increase of the base density. In addition, the corona is denser in the solution for
Case C as compared to Case A. While many of the tangled field lines in Cases A and B are open due to the strong radial
stretching by the stellar wind, in Case C, most of the tangled field lines are closed. The closed loops
in the low corona are radially stretched in Cases A and B, while the same loops are more potential and less stretched
in Case C as seen in the bottom panel of Figure~\ref{fig:f3}.

The interplay between the radial speed and the coronal density structure determines the stellar mass loss rate, 
as well as the stellar angular momentum loss rate to the stellar wind. We follow the method by \cite{Cohen09b}, 
and calculate these loss rates from the MHD solution. This method expands the idealized approach by \cite{weberdavis67} 
and uses the fact that the MHD solution provides a realistic, non-idealized Alfv\'en surface, at which the Alfv\'enic 
Mach number, $M_A=u/v_A=1$, where $v_A=B/\sqrt{4\pi\rho}$ is the Alfv\'en speed. Once the Alfv\'en surface 
has been determined, the loss rates can be calculated as:
\begin{equation}
\dot{M}=\int \rho \mathbf{u}\cdot\mathbf{da}_A,
\label{Mdot}
\end{equation}
\begin{equation}
\dot{J}=\frac{3}{2}\int \Omega_\star \sin{\theta}\;r^2_{A} \;\rho \mathbf{u}\cdot\mathbf{da}_A,
\label{Jdot}
\end{equation}
where $r_A$ is the local radius of the Alfv\'en surface, $\mathbf{da}_A$ is a surface element, 
and the integration is done over the realistic Alfv\'en surface. It is worth to mentioning that the 
realistic Alfv\'en surface, at which the magnetic breaking of the stellar wind takes place, is the 
actual source surface, and it does not have a spherical shape as is assumed to have in the 
common use of the potential field approximation.

The mass and angular momentum loss rates for the different test cases are shown in the upper part of 
Table~\ref{table:t2}. For comparison and verification of these results, we have computed similar wind 
models for the solar case. The same numerical method described above was employed for a solar magnetogram 
obtained during the last solar maximum (Carrington Rotation 1958) by the SoHO 
MDI\footnote{{\tt http://sun.stanford.edu/}}. This solar simulation resulted in a mass loss rate of 
$\dot{M}_\odot\approx 2\cdot10^{-14}\;M_\odot\;Yr^{-1}$ and 
an angular momentum loss rate of $\dot{J}\approx 10^{30}\;g\;cm^2\;s^{-2}$ with the use of base density of 
$2\times 10^{8}$~cm$^{-2}$ (same as Case~A). These are similar to canonical solar values, as expected.

Instead, the loss rates from AB Dor are significantly higher than solar. The angular momentum loss 
rate can be 2 orders of magnitude higher, simply due to the much more rapid rotation of AB~Dor. 
For Case A, the mass loss rate for AB~Dor is about a factor of 10 larger than the equivalent solar case. 
This is perhaps slightly surprising since all parameters other than the rotation rate and, to some extent, 
the surface field map are fairly similar to those of the active Sun chosen for the comparison. The most 
conspicuous difference is the factor of 50 in rotation rate and it is worth examining the influence of 
rotation alone in more detail.

\begin{figure*}[h!]
\centering
\includegraphics[width=6.in]{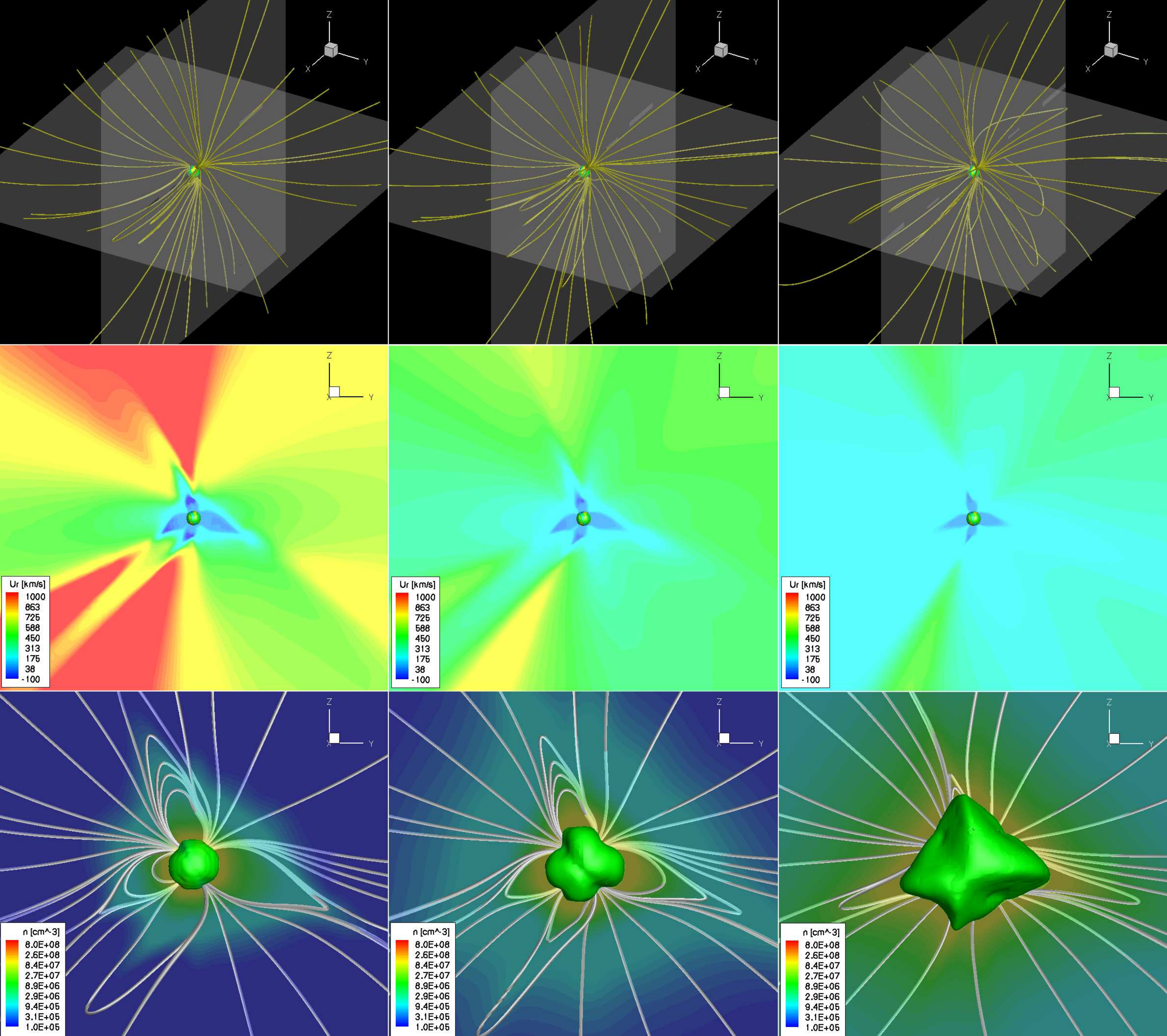}
\caption{Same display as Figure~\ref{fig:f3} but for the cases with $P_\star=25\;d$.}
\label{fig:f4}
\end{figure*}

We repeated the AB~Dor computational runs for Cases A-C for a rotation period of $25\;d$ instead of $0.5\;d$, 
with all other aspects of the simulations remaining the same. The solutions for these runs are shown in 
Figure~\ref{fig:f4}. The azimuthal tangling of the coronal field that characterizes the $0.5\;d$ period 
results in Figure~\ref{fig:f3} is, unsurprisingly, completely absent in this set of solutions and all field 
lines are essentially radial. In addition, more field lines are open in these solutions compared to the case 
with rapid rotation. Mass and angular momentum loss rates are listed at the bottom of Table~\ref{table:t2}: 
mass loss rates are typically a factor of ten lower than for the $0.5\;d$ period results, and, for the Case~A 
base density, are more similar to the solar value. 


\section{DISCUSSION}
\label{sec:Discussion}

The MHD simulations of AB Dor presented here reveals a coronal structure that is manifestly 
different from the well-studied solar corona. These differences are due to the different 
magnetic structure, which is mostly composed of high-latitude, large-scale regions of strong 
magnetic field, and the rapid stellar rotation which induces azimuthal wrapping and tangling 
of magnetic field. This tangling cannot be obtained from the static, non-MHD, potential field 
extrapolation, which is generally useful only for studying the small closed loops near the 
surface where global effects are less important. We note in passing that for stars with 
large-scale regions of strong magnetic field, closed loops are probably significantly larger 
than in the solar case and the choice for the location of source surface location should be at 
greater radial distance than the common use of $R_{ss}=2.5-3.5R_\star$.

The simulation results show that the mass loss and angular momentum loss rates increase with 
increasing coronal base density. The explanation for the former is trivial: introducing a greater 
mass source at the base will necessarily increase the mass flux through a closed surface around 
this source. The latter effect is more subtle and is due to the fact that $\dot{J}\propto\rho u$. 
When increasing the base density, the density drop with height decreases and the volume between 
the stellar surface and the Alfv\'en surface is filled with more mass and as a result, more torque 
is being applied on the rotating star, thus increasing the angular momentum loss.

For a given distribution of $\rho u$, the angular momentum loss rate is directly proportional 
$\Omega_\star$, so it is not surprising that $\dot{J}$ varies with the rotation rate. We find the mass 
loss rate also depends on the stellar rotation rate. This effect is not apparent if the rotation rate 
is not very high. However, the case of the extremely short rotation period of $0.5\;d$ has a significant 
effect. The reason for this behavior can be found in the azimuthal tangling of the coronal field. When 
the rotation is slow, the global topology of the coronal field is radial (as seen in Figure~\ref{fig:f4}). 
In this case, the coronal density profile essentially drops like $r^{-2}$. When strong rotation 
is present, the azimuthal component of both the magnetic field and the flow become important and the 
radial component of the velocity is reduced. The increase in density in the slow wind case is greater 
than the decrease in speed, and as a result, the total value of $\rho u$ increases. Another contributor 
to the angular momentum loss increase with rotation is the shape of the Alfv\'en surface for each case. 

\begin{figure*}[h!]
\centering
\includegraphics[width=6.in]{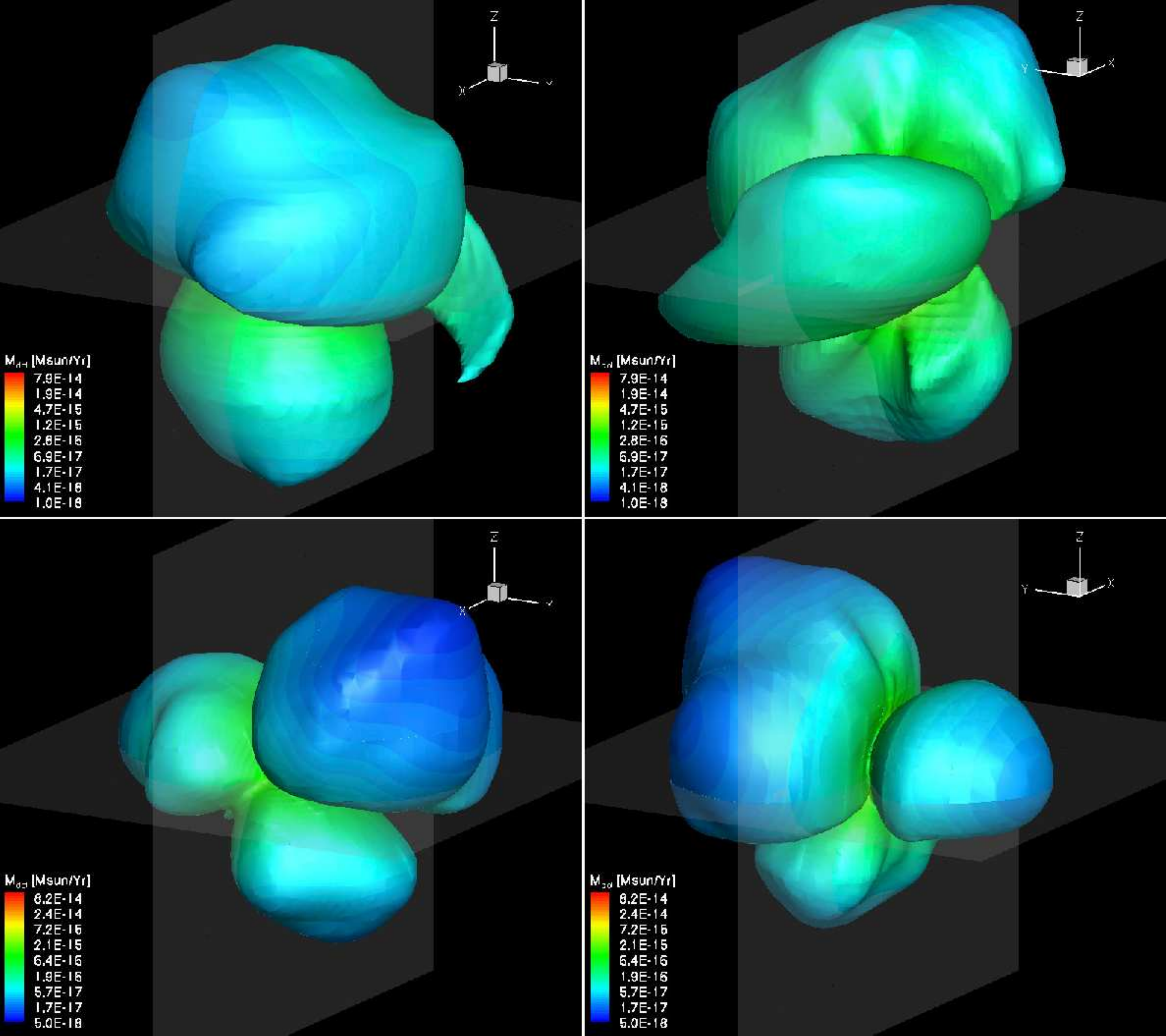}
\caption{The Alfv\'en surface for Case A from different view angle (left-right) colored with contours of the local value of $\dot{M}$ 
with $P_\star=0.5\;d$ (top) and with $P_\star=25\;d$ (bottom). White shades of the $y=0$ and $z=0$ plains are also shown.}
\label{fig:f5}
\end{figure*}

\begin{figure*}[h!]
\centering
\includegraphics[width=6.in]{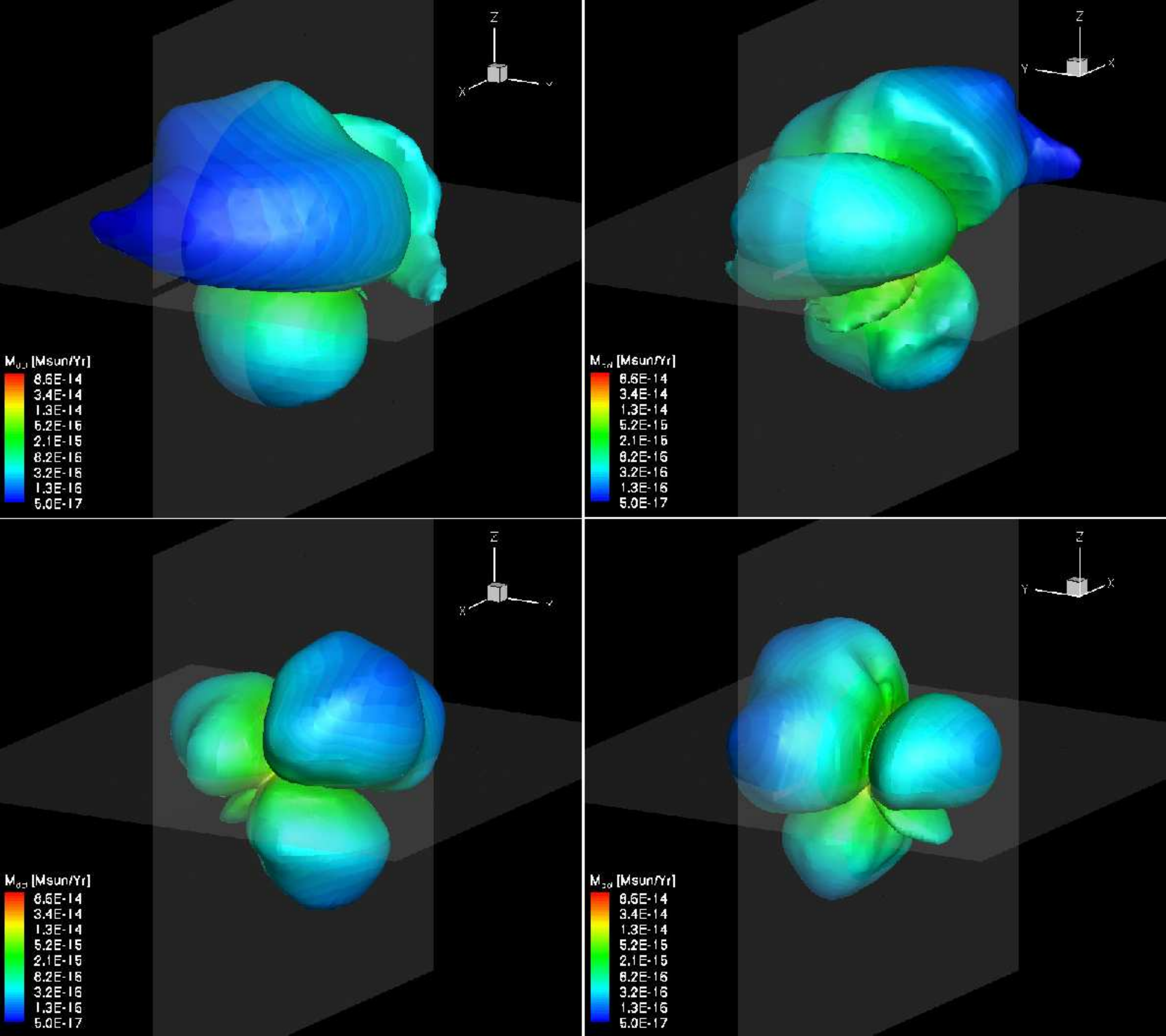}
\caption{Same display as Figure~\ref{fig:f5} but for Case B.}
\label{fig:f6}
\end{figure*}

\begin{figure*}[h!]
\centering
\includegraphics[width=6.in]{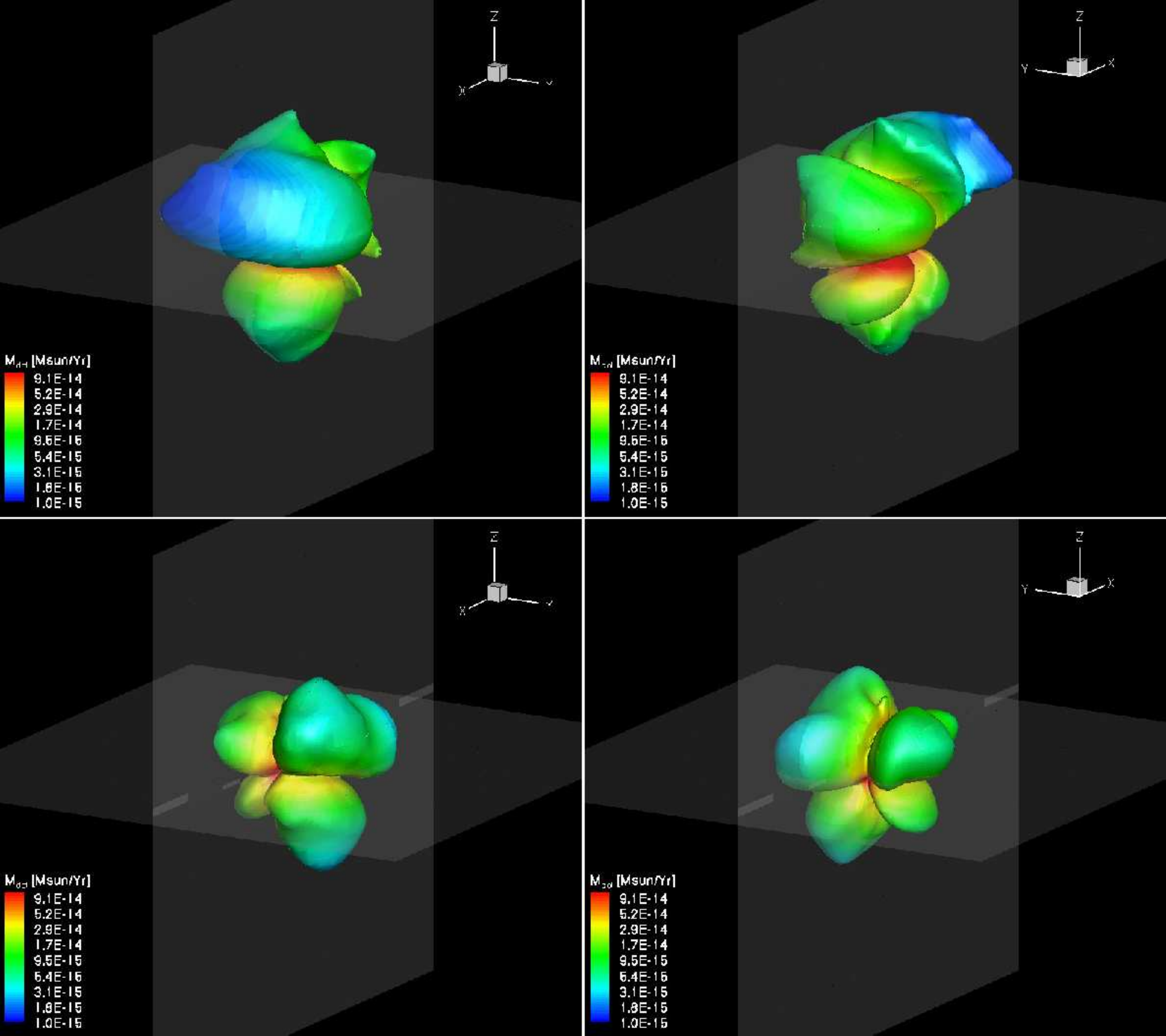}
\caption{Same display as Figure~\ref{fig:f5} but for Case C.}
\label{fig:f7}
\end{figure*}

Figures~\ref{fig:f5}-\ref{fig:f7} show the shape of the Alfv\'en surface for each test case with 
different rotation period. When the rotational period is small, the shape is modified to account for 
the azimuthal component, and the surface is enlarged. In the case of fast 
rotation, the shape of the Alfv\'en surface is modified and signs of the azimuthal component of the 
coronal field can be seen. It also seems like the size of the Alfv\'en surface with faster rotation 
is slightly bigger. The simulations presented here show that the mass loss rate of AB~Dor, 
and presumably also of other rapidly-rotating young late-type stars, is substantially higher than 
the solar mass loss rate, and that it could be as high as $10^{-12}-10^{-11}\;M_\odot\;Yr^{-1}$, as 
suggested by \cite{Wood04,Wood05}. These values however, are strongly dependent on the assumed average 
coronal base density. It seems likely, based on measurements based on X-ray spectra presumably originating 
from plasma in closed loops (Section~\ref{sec:Simulation}) that this is generally higher than the solar case, 
probably by an order of magnitude. For such a case, the predicted mass loss rate for AB~Dor is about 100 
times the solar rate.

The mass and angular momentum loss rates found from our MHD models
here are intriguing for the wider problem of stellar rotational
evolution.  Models such as we present here could, in principle, be
employed to map out theoretical AML as a function of stellar activity
and rotation rate.  While the general picture of stellar spin-down
with age as a result of wind-driven AML, emerged decades ago
\cite[e.g.][]{schatzmann62,weberdavis67,mestel68,skumanich72},
the details of situation has proved somewhat complicated and
rotation rate data amassed in the intervening years for late-type
stars exhibits a complex dispersion over stellar age and mass.

Faster rotation during stellar youth engenders greater magnetic
activity through rotationally-powered dynamo action and the
correlation of magnetic activity indicators such as chromospheric
emission lines and coronal X-ray luminosity with rotation is
well-established.  Observations of Ly$\alpha$ absorption by the
interactions of stellar winds with the surrounding ISM - the stellar
equivalent of the heliopause - also indicate that stellar wind
mass-loss rates are larger for younger and more active stars.
\citet{wood02} estimate a relation $\dot{M}\propto t^{-2.00\pm
0.52}$, based on combining inferred mass-loss rates with X-ray
activity and observed X-ray activity vs stellar age.  Their relation
suggests that at very fast rotation rates, mass-loss should approach
1000 times the solar value, though they caution against the reliability
of this extrapolation.  Here, we find that plausible coronal base
densities lead to mass loss rates of 100 times that of the present day
Sun.

Observations of rotation rates for stars in open clusters indicate
that very young stars with ages of up to 100~Myr or so are not as
rapidly spun-down as would be expected based on extrapolation of the
\citet{skumanich72} spin-down relation, and theoretical spin-down
modelling efforts have invoked a magnetic ``saturation'' that limits
the AML for very short rotation periods
\citep[eg][]{Chaboyer95,krishnamurthi97,barnes03}.  While
our wind model for a single star, such as AB~Dor presented here,
cannot in itself be used to validate such a saturation approach, the
general methodology does in principle allow for a more thorough
exploration of the relevant parameter space to provide an MHD wind
prediction of angular momentum loss as a function of stellar age.  The
requisite input parameters here would be the global coronal base
density, and the magnetic field strength.

The spindown time of AB Dor can be estimated as \citep{UdDoula09}:
\begin{equation}
\tau=\frac{kM_\star\Omega_\star
R^2_\star}{\frac{2}{3}\dot{M}\Omega_\star r^2_A}\approx 0.1\cdot
X^{-2}\dot{M}^{-1}\;[yr], \label{TauSpindown}
\end{equation}
where we used $k=0.1$ and write $r_A=XR_\star$. Based on
Equation~\ref{TauSpindown} and the calculated mass loss rates, the
spindown time of AB Dor can range from $10^9-10^{12}\;yr$ depending on
the particular case and on the value of $X$ (which ranges from 5-10). A
value of $10^9\;yr$ leads to a rotation of $P=0.5\;d\cdot e^5=71\;d$
after 5 billion years. However, \cite{Cohen09b} have shown that the
angular momentum loss rate can be 3-4 times higher when the stellar
magnetic field is dominated by strong polar spots, as appear to
characterize young,
fast-rotating stars. We expect the angular momentum loss rate to
decrease with time as AB Dor becomes an established  main-sequence star with
spots at lower latitudes.  Therefore, after five billion
years a rotation similar to that of the Sun might be expected. In
Equation~\ref{TauSpindown}, $\tau$ is independent of
$\Omega_\star$. Therefore, for the same parameters but with different
rotation rates, we have $P_1/P_2=25\;d/0.5\;d=50$ and
$\dot{J}_2/\dot{J}_1=50$ (the inverse of $\tau_1/\tau_2$). The ratios of
the angular momentum loss rates for Cases A-C with different rotation
rates are 60,68, and 80, respectively. This is consistent with the
expected idealized value.

In principle, we can propose a relation between the wind in our
model and the Alfv\'en radius as follows. Since at the Alfv\'en
radius, $u_{sw}=v_A$ we have:
\begin{equation}
u_{sw}(1/f_s)=v_A(r_A)=\frac{B(r_A)}{\sqrt{4\pi
\rho(r_A)}}=\frac{v_{A0}R^2_\star}{r^2_A}, \label{AlfvenRadius1}
\end{equation}
where we assume that $B(r)=B_0(R_\star/r)^3$ and
$\rho(r)=\rho_0(R_\star/r)^2$, with $B_0$, $\rho_0$, and $v_{A0}$
being the magnetic field, density, and Alfv\'en speed at the flux tube
base, respectively. From Equation~\ref{AlfvenRadius1} we get:
\begin{equation}
\frac{r_A}{R_\star}=\sqrt{\frac{v_{A0}}{u_{sw}(1/f_s)}},
\label{AlfvenRadius2} \end{equation}
which could provide, in principle
the location of the Alfv\'en radius based on a known magnetic field
distribution and surface density. The relation however, is not trivial
due to the non-linear relation between the parameters that define it. In
reality, the radial functions of the magnetic field and density used
in Equation~\ref{AlfvenRadius1} are not necessarily valid within the
Alfv\'en surface. 


\section{SUMMARY AND CONCLUSIONS}
\label{sec:Conclusions}

We have carried out a 3-dimensional global numerical MHD simulation of the corona of AB Dor, 
driven by a Zeeman-Doppler Image magnetic surface map. We studied three test cases with 
different base density and we also compared the solutions with fast and slow stellar rotations. 
We find that the coronal structure of AB Dor is dominated by the azimuthal tangling of the 
coronal magnetic field as a result of rapid rotation. Based on the MHD solution, we calculate 
a realistic Alfv\'en surface, which enables us to estimate the mass and angular momentum loss 
rates. Our main finding is that the mass loss rate is dependent on the value of the average 
coronal base density, as well as the coronal field and stellar wind topology which are affected 
by rapid stellar rotation. The total mass loss rates ranges between 10-500 times the solar mass 
loss rate, while the angular momentum loss rate ranges between 15-30000 times the solar angular 
momentum loss rate. We demonstrate that the global coronal solution depends to some extent on 
the detailed properties of the coronal plasma.

In addition to the uncertainty in stellar parameters, the surface maps used here to drive the model are 
not well defined as well. First, we interpolated the field in the ``missing'' part of the stellar surface, 
second the resolution of the maps in the regions where data is available is not very high, and third, 
the interpretation of these maps is somehow debatable. In particular, large-regions of the 
map appear with strong magnetic field. One can ask whether these are really large-scale strong field 
regions, or whether it is more localized active region that is being smeared by the low resolution. 
The two scenarios should lead to a significantly different MHD solutions.    
 
Further modeling effort like the one presented here should focus on constraining stellar 
parameters such as the stellar wind speed for non solar-like stars. In addition, 
a more consistent model to drive the stellar wind can help to generalize the model 
to stellar systems.


\acknowledgments

We thank for an unknown referee for his/hers useful comments. OC is supported by SHINE through NSF ATM-0823592 grant, and by NASA-LWSTRT Grant NNG05GM44G.
JJD and VLK were funded by NASA contract NAS8-39073 to the {\it Chandra X-ray Center}.
Simulation results were obtained using the Space Weather Modeling
Framework, developed by the Center for Space Environment Modeling, at the University of Michigan with funding
support from NASA ESS, NASA ESTO-CT, NSF KDI, and DoD MURI.



\clearpage


\begin{table}[h!]
\caption{Adopted Properties of AB~Dor}
\begin{tabular}{c||c}
\hline
$n_0$ & $2\cdot 10^8$, $10^9$, $10^{10}\;cm^{-3}$\\
$T_0$ & $5\;MK$\\
$R_\star$ & $0.86R_\odot$ \\
$M_\star$ & $0.76M_\odot$ \\
$P_{rot}$ & $0.5\;d$ \\
\hline
\end{tabular}
\label{table:t1}
\end{table}

\begin{table}[h!]
\caption{Mass and Angular Momentum Loss Rates for AB~Dor.}
\begin{tabular}{l||lll}
\hline
\multicolumn{4}{|c|}{$P_\star=0.5\;d$} \\
\hline
{\bf Test Case} & A & B & C \\
$\dot{M}\;[M_\odot\;Yr^{-1}]$ & $4.5\cdot10^{-13}$ & $2.1\cdot10^{-12}$ & $1.1\cdot10^{-11}$\\ 
$\dot{J}\;[g\;cm^2\;s^{-2}]$ & $2.6\cdot10^{33}$ & $6.8\cdot10^{33}$ & $3.2\cdot10^{34}$ \\
\hline
\multicolumn{4}{|c|}{$P_\star=25\;d$} \\
\hline
{\bf Test Case} & A & B & C \\
$\dot{M}\;[M_\odot\;Yr^{-1}]$ & $4.3\cdot10^{-14}$ & $1.8\cdot10^{-13}$ & $1.6\cdot10^{-12}$\\ 
$\dot{J}\;[g\;cm^2\;s^{-2}]$ & $4.4\cdot10^{31}$ & $1.0\cdot10^{32}$ & $4.4\cdot10^{32}$ \\
\hline
\end{tabular}
\label{table:t2}
\end{table}


\clearpage

\clearpage

\clearpage

\clearpage

\clearpage

\end{document}